\documentclass[paper, nofootinbib]{revtex4}
\usepackage{amsfonts}
\usepackage{graphicx}
\usepackage[brazil]{babel}
\usepackage[latin1]{inputenc}
\usepackage{amsmath}
\usepackage{amssymb}
\begin{document}


\title{Nonminimal spin-field interaction of the classical electron and quantization of spin.}

\author{Alexei A. Deriglazov }
\email{alexei.deriglazov@ufjf.edu.br} \affiliation{Depto. de Matem\'atica, ICE, Universidade Federal de Juiz de Fora,
MG, Brazil,} \affiliation{Department of Physics, Tomsk State University, Lenin Prospekt 36, 634050, Tomsk, Russia}


\begin{abstract}
I shortly describe semi-classical models of spinning electron and list a number of theoretical issues where these models turn out to be useful. Then I discuss the possibility to extend the range of applicability of these models by introducing an interaction, that forces the spin to align up or down relative to its precession axis.
\end{abstract}

\maketitle 

The notion of spin of an elementary particle \cite{Gousmith_1926, Frenkel2, Frenkel1, Thomas1927, Pauli_1927} was developed in attempts to explain the energy levels of
atomic spectra. This analysis culminated in quantum-mechanical expression for the energy of an electron, known as the Pauli Hamiltonian (${\bf E}$ is the Coulomb electric field and ${\bf B}$ is a constant magnetic field, for the vector and scalar product of three-dimensional vectors we use the notation $[{\bf A}, {\bf B}]$ and $({\bf A}, {\bf B})$)
\begin{eqnarray}\label{qs1}
H=\frac{1}{2m}(\hat{\bf p}-\frac{e}{c}{\bf A})^2+eA^0-\frac{e}{mc}\left[(\hat{\bf S}, {\bf B})+\frac{1}{2mc}(\hat{\bf S}, [{\bf E},
\hat {\bf p}])\right].
\end{eqnarray}
Besides the operators of position and momentum, the Hamiltonian contains operators proportional to the Pauli matrices, called the spin operators:  $\hat S^i=\frac{\hbar}{2}\sigma^i$. They have discrete spectrum of eigenvalues, so their contribution to the energy turns out to be quantized. For instance, the operator $(\hat{\bf S}, {\bf B})$ has eigenvalues $\pm\frac{\hbar}{2}|{\bf B}|$, that is the measurement of spin in the direction of vector ${\bf B}$ always gives one of the values $\pm\frac{\hbar}{2}$. 
The extra degrees of freedom contribute to the energy levels of an electron in a good agreement with experiments. The fine structure of hydrogen-like atoms with one valence electron fixes the factor in front of S-E-p term, while Zeeman effect fixes the factor in front of S-B term. 

According the canonical quantization paradigm, the classical analogy of the theory (\ref{qs1}) could be a point particle $({\bf x}, {\bf p})$ which carries a vector ${\bf S}$ attached to it. The commutator $[\hat S^i, \hat S^j]=i\hbar\epsilon^{ijk}\hat S^k$ implies the use of classical-mechanics bracket $\{S^i, S^j\}=\epsilon^{ijk}S^k$. Then the Hamiltonian equation for spin is $d{\bf S}/dt=\{{\bf S}, H\}$, or 
\begin{eqnarray}\label{qs2}
\frac{d{\bf S}}{dt}=[{\bf R}, {\bf S}], \quad \mbox{where} \quad 
{\bf R}\equiv-\frac{e}{mc}\left\{{\bf B}-\frac{1}{2mc}[{\bf p}, {\bf E}]\right\}.
\end{eqnarray}
When ${\bf R}$ is a constant vector\footnote{Consider either frozen particle in constant magnetic field, or on circular trajectory in the Coulomb electric and constant magnetic fields.}, the spin precesses around ${\bf R}$: the end of ${\bf S}$ always lies in a plane orthogonal to ${\bf R}$, and describe a circle around ${\bf R}$ with an angular velocity equal to the magnitude $|{\bf R}|$ of this vector. This, in essence, represents the classical model of non relativistic spin. The model can be constructed on the base of a variational problem that implies both dynamical equations and the magnitude-of-spin constraint ${\bf S}^2=\hbar^2s(s+1)=\frac{3\hbar^2}{4}$, see \cite{AAD_Rec}.  

The relativistic generalization and systematic construction of the resulting theory on the base of a variational problem presents an issue with almost a centenary of  history, see the works \cite{corben:1968, hanson1974, AAD_Rec} and references therein. In the pioneer works \cite{Frenkel2, Frenkel1}, Frenkel identified the components $S^i$ of three-dimensional spin with spatial part of four-dimensional antisymmetric spin-tensor $S^{\mu\nu}=-S^{\nu\mu}$ as follows: $S^{i}=\frac{1}{4}\epsilon^{ijk}S^{jk}$, and 
assumed that at each instant of motion $S^{\mu\nu}$ obeys the covariant condition $S^{\mu\nu}\dot x_\nu=0$, which guarantees the equal number of spin degrees of freedom in relativistic and Pauli theories. Concerning the corresponding variational problem, we recall that the classical-mechanics formalism always implies the canonical Poisson brackets, so the variables with brackets $\{S^i, S^j\}=\epsilon^{ijk}S^k$ are not appropriate to this aim. One possibility to avoid the problem consist in formulating the variational problem in terms of a vector-like basic variable $\omega^\mu$ for the description of spin. Then spatial components of the phase-space quantity $S^{\mu\nu}=2(\omega^\mu\pi^\nu-\omega^\nu\pi^\mu)$, where $\pi_\mu=\frac{\partial L}{\partial\dot\omega^\mu}$, obey the desired brackets as a consequence of canonical brackets $\{\omega^\mu, \pi^\nu\}=\eta^{\mu\nu}$ for the basic variables. This leads to the vector model of a relativistic spin: using $\omega^\mu$ and the square root construction\footnote{The construction can be resumed as follows: variational problem with the 
Lagrangian $L=\dot{\bf x}^2+\dot{\boldsymbol\omega}^2-\sqrt{(\dot{\bf x}^2+\dot{\boldsymbol\omega}^2)^2-4(\dot{\bf x}\dot{\boldsymbol\omega})^2}$ implies the constraint $({\bf p}, {\boldsymbol \pi})\equiv(\frac{\partial L}{\partial\dot{\bf x}}, \frac{\partial L}{\partial\dot{\boldsymbol\omega}})=0$ as one of the extreme conditions.} first discovered by Hanson and Regge \cite{hanson1974}, we can write the following Lagrangian action  \cite{AAD_2014}:
\begin{eqnarray}\label{qs3}
S =-\frac{1}{\sqrt{2}} \int d\tau \left[ m^2c^2 -\frac{\alpha}{\omega^2} \right]^{\frac 12}
\sqrt{-\dot x N \dot x -
\dot\omega N \dot\omega + \sqrt{[\dot x N\dot x + \dot\omega N \dot\omega]^2- 4 (\dot x N \dot\omega )^2}},
\end{eqnarray}
where $N_{\mu\nu}\equiv  \eta_{\mu\nu}-\frac{\omega_\mu \omega_\nu}{\omega^2}$ is a projector on the plane orthogonal to $\omega$. The parameter $\alpha$ determines the value of spin, in particular,
$\alpha=\frac{3\hbar^2}{4}$ corresponds to the spin one-half particle. In the spinless limit, $\omega^\mu=0$ and
$\alpha=0$, Eq. (\ref{qs3}) reduces to the standard Lagrangian of a point particle, $-mc\sqrt{-\dot x^2}$.

The variables $\omega$ and $\pi$ are affected by local symmetry transformations presented in the theory, so they are not an observable quantities. The observable quantities of spin-sector are contained among components of spin-tensor. It obeys the constraints $S^{\mu\nu}S_{\mu\nu}=8\alpha$ and $S^{\mu\nu}p_\nu=0$ which, together with dynamical equations, arise as the conditions of extreme of the variational problem (\ref{qs3}). 

In what follows, we list a number of applications of the vector model. 

{\bf 1. Spin-induced noncommutativity.} An interesting property of classical spin models is that even in non interacting theory the relativistic invariance inevitably leads to non canonical Poisson brackets of dynamical variables. Without going into technical details \cite{DPM2016, AAD_Rec}, this can be explained as follows. The spin supplementary condition $S^{\mu\nu}p_\nu=0$ must be satisfied at all instants of time, hence the equation $\frac{d}{d\tau}(S^{\mu\nu}p_\nu)=0$ should be a consequence of dynamical equations. In the Hamiltonian formulation, variation rate
of a phase-space function is equal to the bracket of this function with Hamiltonian, so we can write
$\frac{d}{d\tau}(S^{\mu\nu}p_\nu)=\{S^{\mu\nu}p_\nu, H\}=\{S^{\mu\nu}p_\nu, z^A\}\frac{\partial H}{\partial z^A}=0$, where $z^A$ is one of $x$, $p$, or $S$. We want to be able to work with different Hamiltonians, so we require\footnote{In the language of Poisson geometry this equation means that $S^{\mu\nu}p_\nu$ are Casimir functions of the Poisson structure.}:
$\{z^A, S^{\mu\nu}p_\nu\}=0$. In particular, $\{x^\alpha, S^{\mu\nu}p_\nu\}=0$ together with $\{x^\alpha, p_\nu\}=\delta^\alpha{}_\nu$ imply nonvanishing bracket of $x^\alpha$ with $S^{\mu\nu}$. The Jacobi 
identity $\{S^{\mu\nu}, \{x^\alpha, x^\beta\}\}+cycle=0$ then implies the nonvanishing brackets of positions $\{x^\mu, x^\nu\}=-\frac{1}{2p^2}S^{\mu\nu}$. For the spatial components this implies
\begin{eqnarray}\label{qs4}
\{x^i, x^j\}=\frac{1}{(mc)^2}\epsilon^{ijk}S^k, \quad i, j=1, 2, 3. 
\end{eqnarray}
The r.h.s. vanishes as $c\rightarrow\infty$, that is the spin-induced noncommutativity is a relativistic effect. It should be noted that spinning particles represent an exceptional example of intrinsically noncommutative and relativistic-invariant theory, with the spin-induced noncommutativity that manifests itself already at the Compton scale. One of immediate consequences of Eq. (\ref{qs4}) is that in $1/c^2$\,-approximation the position of a particle in the Pauli (and Dirac) quantum mechanics should be described by the Pryce (d) operator (see \cite{pryce1948mass, DPM2, Choi_2018} for the details)
\begin{eqnarray}\label{qs5}
\hat x^i=x^i-\frac{\hbar}{4(mc)^2}\epsilon^{ijk}\hat p^j\sigma^k, 
\end{eqnarray}
instead of $\hat x^i=x^i$. 

{\bf 2. The problem of covariant formalism and the Thomas precession.} The Lagrangian (\ref{qs3}) admits interaction with an arbitrary electromagnetic field, and thus gives the relativistic generalization of an approximate Frenkel equations. This leads to the Hamiltonian of interacting theory that has a simple and expected form 
\begin{eqnarray}
H=\frac{1}{2m}\left[(p^\mu-\frac{e}{c}A^\mu)^2-\frac{e\mu}{2c}F_{\mu\nu}S^{\mu\nu}+(mc)^2\right]\approx \qquad  \label{qs6.0}\\
mc^2+\frac{1}{2m}({\bf p}-\frac{e}{c}{\bf A})^2+eA^0-\frac{e}{mc}\left[({\bf S}, {\bf B})+\frac{1}{mc}({\bf S}, [{\bf
E}, {\bf p}])\right], \label{qs6}
\end{eqnarray}
where in the last line we put the magnetic moment $\mu=1$, and left $1/c^2$ approximation of the complete Hamiltonian in the physical-time parametrization. It is accompanied by higher nonlinear Poisson brackets, that is the most part of interaction in the vector model is encoded in the noncommutative phase-space geometry. The relativistic (\ref{qs6}) and Pauli (\ref{qs1}) Hamiltonians differ by the famous $1/2$ factor in front of the last terms. This is the problem of covariant formalism, that was raised already in the pioneer works \cite{Gousmith_1926, Frenkel2, Thomas1927} and remains under debates to date. The spin-induced noncommutativity of vector model provides a natural solution to this problem: the Hamiltonian (\ref{qs6})  is accompanied by non canonical brackets, and this should be taken into account during the quantization.  Detailed computations shows \cite{DPM2016} that this results in Pauli's quantum mechanics.

It should be noted that this result is obtained without any appeal to the effect of Thomas precession. The role of the Thomas spin-vector in this scheme was clarified in the recent work \cite{Dan2019}. 

{\bf 3. Modified Mathisson-Papapetrou-Tulczyjew-Dixon (MPTD) equations.} Vector model admits the minimal interaction with an arbitrary gravitational field, it is sufficient to replace $\eta_{\mu\nu}$ on $g_{\mu\nu}$, and $\dot\omega^\mu$ on $\nabla\omega^\mu=\dot \omega^\mu+\Gamma^{\mu}{}_{\alpha\beta}\dot x^\alpha\omega^\beta$ in the Lagrangian (\ref{qs3}). Remarkably, this leads to equations of motion that coincide with MPTD equations in the form studied by Dixon \cite{Dixon1964}.  So the vector model gives an alternative description of a rotating body in gravitational field in the dipole approximation. 
In particular, using the vector model, it is possible to compute and analyse the three-dimensional acceleration of MPTD body. This analysis shows that MPTD equations lead to a wrong dependence of acceleration on speed of the body. 

Without going into technical details \cite{AAD_Rec}, this can be explained as follows. In special and general relativity, the acceleration of a particle inevitably depends on its velocity in such a way that longitudinal acceleration vanishes as $v\rightarrow c$. In electrodynamics we have
\begin{eqnarray}\label{qs7}
\frac{dv^\mu}{ds}=F^\mu{}_\nu v^\nu, \quad \mbox{implies} \quad {\bf a}_{||}\sim(c^2-{\bf v}^2)^{\frac32}\stackrel{v\rightarrow c}{\longrightarrow}0,
\end{eqnarray}
while for the geodesic equation
\begin{eqnarray}\label{qs8}
\frac{dv^\mu}{ds}=-\Gamma^\mu{}_{\nu\alpha} v^\nu v^\alpha \quad \mbox{implies} \quad {\bf a}_{||}\sim (c^2-{\bf v}^2) \stackrel{v\rightarrow c}{\longrightarrow}0. 
\end{eqnarray}
These examples illustrate the general rule: each four-velocity $v^\mu$, appeared on r. h. s. of manifestly covariant equation in proper-time parametrization, contributes the factor $(c^2-{\bf v}^2)^{-\frac12}$ into the expression for ${\bf a}_{||}$. 
So the relativistic equations of motion can not contain too many velocities $v^\mu$ (four or more).  Unfortunately, MPTD equations belong to this last case \cite{AAD_Rec}. In the result,  acceleration grows with speed and diverges in the ultra-relativistic limit  $v\rightarrow c$. Therefore, MPTD equations do not seem to be a promising candidate for describing a rotating body. The particles/bodies with spin are now under intensive investigation and represent an important tool in the study of near horizon physics of black holes, see for example \cite{Nica_2019, Tosh_2019, Kay_2019, Saj_2019, Plya_2019}. So, it is interesting to find a generalization of MPTD equations with an improved dependence of acceleration on speed. This can be achieved by adding a nonminimal spin-gravity  interaction \cite{AAD_Rec}. In the Hamiltonian formulation, this reduces to the replacement of original minimally-interacting Hamiltonian ${\cal P}^2+(mc)^2$ on the following: ${\cal P}^2+\kappa R_{\mu\nu\alpha\beta}S^{\mu\nu}S^{\alpha\beta}+(mc)^2$. The modified Hamiltonian strongly resembles to that of spinning particle with a magnetic moment (\ref{qs6.0}), so the coupling constant $\kappa$ is called gravimagnetic moment \cite{Khriplovich1989}. $\kappa=0$ corresponds to the MPTD equations. The
most interesting case turns out to be $\kappa=1$ (gravimagnetic body). Keeping only the terms, which may contribute in
the leading post-Newtonian approximation, this gives the modified equations
\begin{eqnarray}\label{qs9}
\nabla P_\mu = -\frac 14 \theta_{\mu\nu}\dot x^\nu -\frac{\sqrt{-\dot x^2}}{32mc} (\nabla_\mu  \theta_{\sigma\lambda})
S^{\sigma\lambda}, \quad  \nabla S^{\mu\nu} =\frac{\sqrt{-\dot x^2}}{4mc}\theta^{[\mu}{}_{\alpha} S^{\nu]\alpha}, 
\qquad \qquad
\end{eqnarray}
where $\theta_{\mu\nu}=R_{\mu\nu\alpha\beta}S^{\alpha\beta}$ is a gravitational analogy of the electromagnetic field
strength $F_{\mu\nu}$.  They can be compared with MPTD equations in the same approximation
\begin{eqnarray}\label{qs10}
\nabla P_\mu =-\frac{1}{4}\theta_{\mu\nu}\dot
x^\nu,  \qquad
\nabla S^{\mu\nu} = 0.
\end{eqnarray}
We see that unit gravimagnetic moment yields quadratic in spin corrections to MPTD equations in the $1/c^2$\,-approximation. 

Both acceleration and spin torque of gravimagnetic body have a reasonable behaviour in the modified theory \cite{AAD_Rec}.  In Schwarzschild and Kerr space-times, the modified equations predict a number of qualitatively new effects \cite{WAD_1917}, that could be used to test experimentally, whether a rotating body in general relativity has null or unit gravimagnetic moment.

{\bf 4. Dirac equation and unobservability of Zitterbewegung.} Consider the Dirac equation
$i\hbar\partial_t\Psi=\hat H\Psi$, where $\hat H= c\alpha^i\hat
p_i+mc^2\beta$,
and  $\alpha^i=\gamma^0\gamma^i$ and $\beta=\gamma^0$ are Dirac matrices. Passing from the Schr\"odinger to Heisenberg picture, the time
derivative of an operator $a$ is $i\hbar\dot a=[a, H]$, and for the expectation values of basic operators of the Dirac
theory we obtain the equations
\begin{eqnarray}\label{qs12}
\dot x_i=c\alpha_i, \qquad \quad \qquad \dot p_i=0,  \qquad \qquad \cr
i\hbar\dot\alpha_i=2(cp_i-H\alpha_i), \quad  i\hbar\dot\beta=-2c\alpha_ip_i\beta+mc^2.
\end{eqnarray}
They can be solved, with the result for $x^i(t)$ being
$x^i=a^i+dp^it+c^i\mbox{exp}(-\frac{2iH}{\hbar}t)$. The first and second terms are expected, and describe a motion along
the straight line. The last term on the r.h.s. of this expression states that the free electron experiences rapid
oscillations with higher frequency $\frac{2H}{\hbar}\sim\frac{2mc^2}{\hbar}$, called the Zitterbewegung.

Although it is widely believed that Zitterbewegung could be a physically observable phenomenon, a detailed analysis does not support this belief \cite{ZitD_2012, Sil_2019}. The exact relation between the variable of position of vector model and the Dirac operators was computed in \cite{DPM2}. It is different from the naive expressions implied by Eq. (\ref{qs12}), and without going into technical details can be described as follows. A long time ago Feynman noticed (see p. 48 in \cite{Feyn61}), that the basic Dirac operators can be used to construct an operator that do not experiences Zitterbewegung. It is just Pryce (d) operator. But as we saw above, in the vector model namely the  Pryce (d) operator represents the classical variable of position (the latter moves along a straight line according to classical equations). The same conclusion follows from the analysis of one-half of the Dirac equation in the Foldy-Wouthuysen representation \cite{Sil_2019, Sil_2016}.

{\bf 5. Relativistic quantum mechanics with positive energies and the Dirac equation.} Canonical quantization of the vector model (\ref{qs3}) leads to the Schr\"{o}dinger equation
\begin{equation}\label{qs13}
i\hbar\frac{d\Psi}{dt}=c\sqrt{\hat{{\bf p}}^{\,2}+(mc)^2}\Psi, \qquad \hat p^i=-i\hbar\frac{\partial}{\partial x^i}, 
\end{equation}
on the space of wave functions which are two-component Weyl spinors $\Psi(t, {\bf x})=(\Psi_1, \Psi_2)$, in the representation of the operator $\hat x^i=x^i$ conjugated to $\hat p^i=-i\hbar\frac{\partial}{\partial x^i}$.   The scalar product is  
\begin{equation}\label{qs14}
\langle \Psi,\Phi \rangle=\int d^3x\Psi^\dagger \Phi, 
\end{equation}
and $P=\Psi^\dagger \Psi$, is a probability density of $\hat x^i$. All solutions to the Schr\"{o}dinger equation form the subspace of positive-energy solutions to the (again two-component) Klein-Gordon equation
\begin{equation}\label{qs15.1}
(\hat p^2+m^2c^2)\Psi\equiv \sigma^\mu \hat p_\mu\bar\sigma^\nu\hat p_\nu \psi+m^2c^2\psi=0.
\end{equation}
The novel point is that the operators $\hat x^i=x^i$ and $\sigma^i$ do not represent operators of position and spin in the vector model. As we saw above, the classical variables $x^i$ and $S^i$ obey noncanonical brackets, so the corresponding operators 
\begin{equation}\label{qs16}
x^i ~  \rightarrow ~ \hat{X}^{i}=x^i+\frac{\hbar}{2mc(\hat p^0+mc)}\epsilon^{ijk}\sigma_j\hat{p}_k,
\end{equation}
\begin{equation}\label{qs17}
\hat S^i=\frac{\hbar}{2mc}\left(\hat p^0\sigma^i-\frac{1}{(\hat
p^0+mc)}(\hat{\bf{p}}{\boldsymbol{\sigma}})\hat p^i\right),
\end{equation}
turn out to be Pryce (d) operators. 

To complete the construction, it remains to show the relativistic covariance of the quantum-mechanical formalism. Here we only outline the proof of relativistic invariance of the scalar product (\ref{qs14}). Concerning the covariant rules for computation of transition probabilities and mean values of operators, see \cite{DPM2, AAD_Rec}.  Introduce the following operator (we denote $\sigma^\mu=({\bf 1}, \sigma^i)$,  and $\bar\sigma^\mu=(-{\bf 1}, \sigma^i)$):
\begin{equation}\label{qs18}
V=\frac{1}{mc}\sqrt{\frac{\hat p^0}{\hat p^0+mc}}[(\bar \sigma \hat p)+mc], \quad  V^{-1}=\frac{1}{2\sqrt{\hat p^0(\hat
p^0+mc)}}[mc-\sigma\hat p],
\end{equation}
commuting with the Schr\"{o}dinger operator (\ref{qs13}), then the 
vector $\psi=V^{-1}\Psi$ obeys (\ref{qs13}) and (\ref{qs15.1}) together with $\Psi$. Then we can write
\begin{eqnarray}\label{qs19}
\langle \Psi,\Phi \rangle=\langle V\psi,
V\phi \rangle=\int d^3x \frac{1}{m^2c^2}(\bar\sigma\hat p\psi)^\dagger\bar\sigma\hat p\phi+\psi^\dagger\bar\phi
\equiv \int d^3x I^0.
\end{eqnarray}
The integrand is the null component of a four-vector
\begin{eqnarray}\label{qs20}
I^\mu[\psi,\phi]=\frac{1}{m^2c^2}(\bar\sigma\hat p\psi)^\dagger\sigma^\mu\bar\sigma\hat p\phi- \psi^\dagger\bar\sigma^\mu\phi\,,
\end{eqnarray}
that represents a conserved current of Eq. (\ref{qs15.1}), that is $\partial_\mu I^{\mu}=0$, when $\psi$ and $\phi$ satisfy to
Eq. (\ref{qs15.1}). So we can construct a scalar product using the invariant integral: $(\psi,\phi)\equiv\int\limits_{\Omega} d\Omega_\mu I^\mu$, $d\Omega_\mu=-\frac16\epsilon_{\mu\nu\alpha\beta}dx^\nu dx^\alpha dx^\beta$, 
computed over a space-like three-surface $\Omega$. Using the Gauss theorem for the four-volume contained between the
surfaces $\Omega_1$ and $\Omega_2$, we conclude that the scalar product does not depend on the choice of the surface,
$\int_{\Omega_1}=\int_{\Omega_2}$. In particular, it does not depend on time. So we can restrict ourselves to the
hyperplane defined by the equation $x^0=\mbox{const}$, this reduces $(\psi,\phi)$ to the expression 
written in Eq. (\ref{qs19}): $(\psi,\phi)=\int d^3x I^0$. 
Besides, the scalar product is positive-defined, since $I^0[\psi,\psi] > 0$.
In the result, Eq. (\ref{qs14}) represents the relativistic invariant scalar product. 

The operator (\ref{qs18}) is closely related with the Foldy-Wouthuysen transformation of the Dirac equation. To see this, we use the equivalence between Klein-Gordon and Dirac equations contained in the map
\begin{eqnarray}\label{cq22.1}
\Psi_D[\psi]=\frac{1}{\sqrt{2}}\left(
\begin{array}{cc}
\,\,\,\,{\bf 1} & {\bf 1}\\
-{\bf 1} & {\bf 1}
\end{array}
\right)\left(
\begin{array}{c}
\psi\\
\frac{1}{mc}(\bar\sigma\hat p)\psi
\end{array}
\right)=\frac{1}{\sqrt{2}mc}\left(
\begin{array}{c}
{[(\bar\sigma\hat p)+mc]}\psi\\
{[(\bar\sigma\hat p)-mc]}\psi
\end{array}
\right),
\end{eqnarray}
that relates solutions $\psi$ of Klein-Gordon equation (\ref{qs15.1}) with the solutions $\Psi_D$ of the Dirac equation $(\gamma^\mu\hat p_\mu+mc)\Psi_{D}=0$, where $\gamma^\mu$ are $\gamma$\,-matrices in the Dirac representation. Applying the Foldy-Wouthuysen transformation
\begin{eqnarray}\label{cq28}
U_{FW}=\frac{\omega_p+mc+(\vec{\gamma}\vec{p})}{\sqrt{2(\omega_p+mc)\omega_p}},
\end{eqnarray}
to the  Dirac spinor $\Psi_D[\psi]$, we
obtain
\begin{eqnarray}
U_{FW}\Psi_D[\psi]=
\left(
\begin{array}{c}
V\psi\\
0
\end{array}
\right)=\left(
\begin{array}{c}
\Psi\\
0
\end{array}
\right),
\end{eqnarray}
that is the operator $V$ is a restriction of $U_{FW}$ to the space of positive-energy right Weyl spinors $\psi$.

We emphasize that in this paragraph we did not try to give a physical interpretation of manifestly covariant Klein-Gordon and Dirac equations (the interpretation of negative-energy states and so on). Vector model leads to the relativistic quantum mechanics with positive-energy states, which is defined by the expressions (\ref{qs13})-(\ref{qs17}). The covariant Klein-Gordon and Dirac formalisms were used as an auxiliary tool to test the relativistic invariance of this theory.

{\bf Modified equation for precession of spin.} As we saw above, the classical models of spinning electron clarify a number of theoretical issues. However, due to the absence of a classical mechanism similar to the quantization of angular momentum and spin  and the corresponding rules for the addition of moments, the range of applicability of these models for practical calculations is rather limited. For instance, as the basic motion of spin in magnetic field is a precession, its contribution into the classical energy is $({\bf S}, {\bf B})\sim\cos\theta_0$, and can be an arbitrary number depending only on the initial value of the angle 
$0<\theta_0<\pi$ between magnetic field and spin. Let us discuss a possibility to modify the basic motion,  introducing a nonminimal interaction that causes the spin to align up or down relative to its precession axis.

Introduce the matrices $N^{ij}(S)\equiv\delta^{ij}-\frac{S^i S^j}{{\bf S}^2}\equiv \delta^{ij}-\hat S^i \hat S^j$ and $N^{ij}(B)\equiv\delta^{ij}-\frac{B^iB^j}{{\bf B}^2}\equiv\delta^{ij}-\hat B^i \hat B^j$, which are projectors on the plane orthogonal to the unit vectors $\hat{\bf S}=\frac{{\bf S}}{|{\bf S}|}$ and $\hat{\bf B}=\frac{{\bf B}}{|{\bf B}|}$. 
\begin{figure}[t] \centering
\includegraphics[width=210pt, height=150pt]{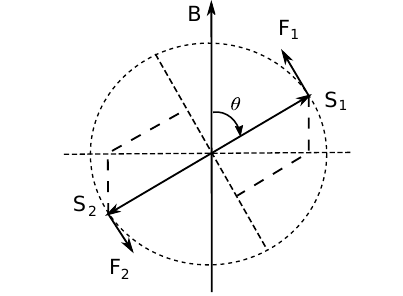}
\caption{Vector of force ${\bf F}_i\sim -N(S_i)N(B){\bf S_i}$ always is directed towards the straight line of the vector ${\bf B}$.}\label{QS-1}
\end{figure}
Using them, we consider the following double projection of ${\bf S}$ (see Fig. \ref{QS-1}): 
\begin{eqnarray}\label{qss1}
-|{\bf B}|N(S)N(B){\bf S}=({\bf B}, {\bf S})N(S)\hat{\bf B}=({\bf B}, {\bf S})[\hat{\bf S}, [\hat{\bf B}, \hat{\bf S}]].
\end{eqnarray}
This vector lies on the plane of ${\bf B}$ and ${\bf S}$, is tangent to the circle of the radius $|{\bf S}|$, and, regardless of the angle between ${\bf B}$ and ${\bf S}$, is always directed towards the straight line of the vector ${\bf B}$. Assuming that a nonminimal interaction of spin with magnetic field is proportional to this vector, we replace the precession equation on the following one: 
\begin{eqnarray}\label{qss1}
\frac{d{\bf S}}{dt}=-\frac{e}{mc}[{\bf B}, {\bf S}]+\beta ({\bf B}, {\bf S})[\hat{\bf S}, [\hat{\bf B}, \hat{\bf S}]],
\end{eqnarray}
where, on dimensional grounds, $\beta=\gamma\frac{|e|}{mc}$, and $\gamma>0$ is a dimensionless constant. As we have a first-order differential equation, the end point of ${\bf S}$ moves along the integral lines of the vector field written on r.h.s. of this equation. So, the evolution of ${\bf S}$ consist of two motions: precession around ${\bf B}$ caused by first term, plus circular motion on the plane of precession (that is on the plane of ${\bf B}$ and ${\bf S}(t)$) caused by second term. Due to the circular motion, a vector of spin that originally had an acute angle with ${\bf B}$ lines up in the direction of ${\bf B}$, while a vector that had an obtuse angle lines up in the opposite to ${\bf B}$ direction. These two directions are stable with respect to small perturbations. The time of alignement depends on the 
angular velocity of the circular motion, which is $\frac{d\theta}{dt}=\frac{\gamma|e{\bf B}\sin 2\theta|}{2mc}$. So the mean time of the alignement is determined by the coupling constant $\gamma$

\vspace{5mm}

{\bf Acknowledgments.} The work of A. A. D. has been supported by the Brazilian foundation CNPq (Conselho Nacional de
Desenvolvimento Cient\'ifico e Tecnol\'ogico - Brasil),  and by Tomsk State University Competitiveness Improvement
Program.

\end{document}